\documentclass[runningheads]{llncs}
\usepackage{amsfonts}
\usepackage{amssymb}
\usepackage[utf8]{inputenc}
\usepackage{listings}
\usepackage{tikz}
\usetikzlibrary{positioning}
\usetikzlibrary{graphs}
\usepackage[hidelinks,colorlinks=true,linkcolor=blue,citecolor=blue,urlcolor=blue]{hyperref}
\usepackage{breakurl}

\newcommand{\Title}{Towards a Verified Model of the Algorand Consensus Protocol in Coq}

\input{lstcoq.sty}

\begin{document}
\title{\Title}
%
%
\author{
  Musab A. Alturki\inst{1} \and
  Jing Chen\inst{2} \and
  Victor Luchangco\inst{2} \and
  Brandon Moore\inst{1} \and
  Karl~Palmskog\inst{3} \and
  Lucas Pe\~na\inst{4} \and
  Grigore Ro\c su\inst{4}
}
\authorrunning{M. A. Alturki et al.}
%
\institute{
  Runtime Verification, Inc., Urbana, IL, USA\\
  \email{\{musab.alturki,brandon.moore\}@runtimeverification.com}\\
  \and
  Algorand, Inc., Boston, MA, USA\\
  \email{\{jing,victor\}@algorand.com}\\
  \and
  KTH Royal Institute of Technology, Stockholm, Sweden\\
  \email{palmskog@acm.org}\\
  \and
  University of Illinois at Urbana-Champaign, Urbana, IL, USA\\
  \email{\{lpena7,grosu\}@illinois.edu}
}
\maketitle              
\begin{abstract}
The Algorand blockchain is a secure and decentralized public ledger
 based on pure proof of stake rather than proof of work.
At its core it is
a novel consensus protocol with exactly one block certified in
each round: that is, the protocol guarantees that the blockchain does not fork.
In this paper, we report on our effort
to model and formally verify the Algorand consensus protocol
in the Coq proof assistant.
Similar to previous consensus protocol verification efforts,
we model the protocol as a state transition system and reason over reachable global states.
However, in contrast to previous work,
our model explicitly incorporates timing issues
(e.g., timeouts and network delays)
and adversarial actions, reflecting a
more realistic environment faced by a public blockchain.

Thus far, we have proved \emph{asynchronous safety} of the protocol:
two different blocks cannot be certified
in the same round,
even when the adversary has complete control
of message delivery in the network.
We believe that our model is sufficiently general
and other relevant properties of the protocol such as liveness
can be proved for the same model.


\end{abstract}

\section{Introduction}
\label{sec:intro}

The Algorand blockchain is a scalable and permissionless public ledger
for secure and decentralized digital currencies and transactions.
To determine the next block,
it uses a novel consensus protocol~\cite{algorand-features-spec,Chen2019} based on pure proof of stake.
In contrast to Bitcoin \cite{Nakamoto2008} and other blockchains based on proof of work,
where safety is achieved by making it computationally expensive to add blocks,
Algorand's consensus protocol is highly efficient and does not require solving cryptographic puzzles.
Instead, it uses \emph{cryptographic self-selection},
which allows each user to individually
determine whether it is selected into the committees responsible for generating the next block.
The self-selection is done independently by every participant,
with probability proportional to its stake.
Private communication channels are not needed;
committees propagate their messages in public.
They reach Byzantine consensus on the next block and certify it, 
so that all users learn the next block without ambiguity.
That is, rather than waiting for a long time so as to be sure that a block will not disappear from the longest chain, as in Bitcoin,
the Algorand blockchain does not fork: a certified block is immediately final, and transactions contained in it can be relied upon right away.
The Algorand blockchain guarantees fast generation of blocks as long as the underlying propagation network is not partitioned
(i.e., as long as messages are delivered in a timely fashion).
The Algorand consensus protocol, its core technology, and mathematical proofs of its safety and liveness properties
are described in \cite{algorand-features-spec,Chen2018,Chen2019}.

The focus of this work is to formally model and verify the Algorand consensus protocol
(described in \cite{Chen2018,Chen2019}) using the Coq proof assistant.
Automated formal verification of desired properties
adds another level of assurance about its correctness,
and developing a precise model to capture
the protocol's runtime environment and the assumptions it depends on
is interesting from a formal-methods perspective as well.
For example, \cite{Woos2016} proves state machine safety and linearizability for the
Raft consensus protocol in a non-Byzantine setting,
and \cite{Pirlea2018} focuses on safety properties of blockchains and,
using a largest-chain-based fork-choice rule and a clique network topology, proves eventual consistency
for~an abstract parameterized protocol.
Similar to previous work,
we define a \mbox{transition} system relation on global protocol states
and reason inductively over \emph{traces} of states
reachable via the relation from some initial state.
We abstract away \mbox{details} on cryptographic primitives,
modeling them as functions with the desired properties.
We also omit details related to blockchain transactions and currencies.

However, our goal and various aspects of the Algorand protocol
present new challenges.
First, our goal is to verify the protocol's asynchronous safety under Byzantine faults.
Thus, we explicitly allow arbitrary adversarial actions, such as user corruption and message replay.
Also, rather than assuming a particular network topology,
the Algorand protocol assumes that messages are delivered within given real-valued deadlines
when the network is not partitioned
(messages may be arbitrarily delayed and their delivery is fully controlled by the adversary
when the network is partitioned).
We capture this by explicitly modeling global time progression and message delivery deadlines in the underlying propagation network.
Moreover, as mentioned above, the Algorand protocol uses cryptographic self-selection to randomly select committees responsible for generating blocks.
As mechanizing probabilistic analysis is still an open field in formal verification,
instead of trying to fully capture randomized committee selection,
we identify properties of the committees
that are used to verify the correctness of the protocol
without reference to the protocol itself.
We then express these properties as axioms in our formal model.
Pen-and-paper proofs that these properties hold
(with overwhelming probability)
can be found in \cite{algorand-features-spec,Chen2019}.

It is worth pointing out that our approach is based on reasoning about \emph{global} states, 
in contrast to \cite{Rahli2018},
which formally verifies the PBFT protocol under\linebreak arbitrary local actions.
While it is possible to model coordinated actions as in~\cite{Rahli2018}, our model
explicitly allows an adversary to arbitrarily coordinate
actions (at the network level) among corrupted users using both newly forged and valid past messages. 
Finally, \cite{Sergey2018} uses distributed separation logic for consensus protocol verification in Coq
with non-Byzantine failures.
Using this approach to verify protocols under Byzantine faults is an interesting avenue of future work.

Thus far, we have proved in Coq \emph{asynchronous safety}:
two different blocks can never be certified
in the same round,
even when the adversary has complete control
of the network.
We believe that our model is sufficiently general
to allow other relevant properties of the protocol such as liveness
to be proved.

\section{The Algorand Consensus Protocol}
\label{sec:protocol}

In this section,
we give a brief overview of the Algorand consensus protocol
with details salient to our formal model.
More details can be found in~\cite{algorand-features-spec,Chen2019,Gilad2017}.

All users participating in the protocol have unique identifiers (public keys).
The protocol proceeds in \emph{rounds} and 
each user learns a 
{\em certified} block
for each round.
Rounds are asynchronous:
each user individually starts a new round
whenever it learns a certified block
for its current round.

A round consists of one or more \emph{periods},
which are attempts to generate a certified block.
A period consists of several \emph{steps}:
users propose blocks
and then vote to certify a proposal.
Specifically,
each user waits a fixed amount of time (determined by network parameters) to receive proposals,
and then votes to support the proposal with the best \emph{credential},
as described below;
these votes are called \emph{soft-votes}.
If it receives a quorum of soft-votes,
it then votes to certify the block;
these votes are called \emph{cert-votes}.
A user considers a block certified
if it receives a quorum of cert-votes.
If a user doesn't receive a quorum of cert-votes
within a certain amount of time,
it votes to begin a new period;
these votes are called \emph{next-votes}.
A next-vote may be for a proposal,
if the user received a quorum of soft-votes for it,
or it may be \emph{open}.
A user begins a new period
when it receives a quorum of next-votes from the same step for the same proposal 
or a quorum of open next-votes;
and repeats the next-vote logic otherwise.

\paragraph{Committees.}
For scalability, not all users send their messages in 
every step.
Instead,
a committee is randomly selected for each step
via a technique called \emph{cryptographic self-selection}:
each user independently determines
whether it is in the committee
using a \textit{verifiable random function} (VRF). 
Only users in the committee
send messages for that step,
along with a \emph{credential} generated by the VRF
to prove they are selected.
Credentials are totally ordered,
and the ones accompanying proposals are used to determine 
which proposal to support.

\paragraph{Network.}
Users communicate by propagating messages over the network.  Message 
delivery is asynchronous and may be out-of-order, but delivery times are bounded: 
any message sent or received by an honest user is received by all honest users within 
a fixed amount of time unless the network is \emph{partitioned}.  (There is no bound on message
delivery time if the network is partitioned.)

\paragraph{Adversary.}
The adversary can corrupt any user and control and coordinate corrupted users' actions:
for example, to resend old messages,
send any message for future steps of the adversary's choice,
and decide when and to whom the messages are sent by them.
The adversary also controls when messages are delivered between honest users
within the bounds described above, and fully controls message delivery when the network is partitioned.
The adversary must control less than 1/3 of the total stake participating in the consensus protocol.

\section{Model}
\label{sec:model}

Our Coq model of the protocol, which is an abstracted version of the latest Algorand consensus 
protocol described in~\cite{Chen2018,Chen2019}, is a transition system
encoded as an inductive binary relation on global states.
The transition relation is parameterized on finite types of
user identifiers (\lstinline[language=Coq,basicstyle=\ttfamily\normalsize]{UserId})
and values (\lstinline[language=Coq,basicstyle=\ttfamily\normalsize]{Value});
the latter abstractly represents blocks and block hashes.

\paragraph{User and Global State.}
We represent both user state and global state as Coq records.
For brevity,
we omit a few components of the user state in this paper
and only show some key ones,
such as the Boolean indicating whether a user is corrupt,
the local time, round, period, step,
and blocks and cert-votes that have been observed.
The global state has the global time,
user states and messages via finite maps~\cite{Cohen2019},
and a Boolean indicating whether the network is partitioned.

\noindent\begin{minipage}[t]{0.45\textwidth}
\begin{lstlisting}[language=Coq]
Record UState := mkUState {
 corrupt: bool; timer: R;
 round: nat; period: nat; step: nat;
 blocks: nat -> seq Value;
 certvotes: nat -> nat -> seq Vote;
 (* ... omitted ... *)
}.
\end{lstlisting}
\end{minipage}
\begin{minipage}[t]{.45\textwidth}
\begin{lstlisting}[language=Coq]
Record GState := mkGState {
 network_partition: bool;
 now: R;
 users: {fmap UserId -> UState};
 msgs: {fmap UserId -> {mset R * Msg}};
 msg_history: {mset Msg};
}.
\end{lstlisting}
\end{minipage}

\paragraph{State Transition System.}
The transition relation on global states \lstinline[language=Coq,basicstyle=\ttfamily\normalsize]{g} and \lstinline[language=Coq,basicstyle=\ttfamily\normalsize]{g'},
written \lstinline[language=Coq,basicstyle=\ttfamily\normalsize]{g ~~> g'},
is defined in the usual way via inductive rules.
For example, the rule for adversary message replay is as follows:
\begin{lstlisting}[language=Coq]
step_replay_msg : forall (pre:GState) uid (ustate_key : uid \in pre.(users)) msg,
  not pre.(users).[ustate_key].(corrupt) -> msg \in pre.(msg_history) ->
  pre ~~> replay_msg_result pre uid msg
\end{lstlisting}
Here, \lstinline[language=Coq,basicstyle=\ttfamily\normalsize]{replay_msg_result} is a function
that builds a global state where \lstinline[language=Coq,basicstyle=\ttfamily\normalsize]{msg} is broadcast.
We call a sequence of global states a \emph{trace}
if it is nonempty
and \lstinline[language=Coq,basicstyle=\ttfamily\normalsize]{g ~~> g'} holds
whenever \lstinline[language=Coq,basicstyle=\ttfamily\normalsize]{g} and \lstinline[language=Coq,basicstyle=\ttfamily\normalsize]{g'} are adjacent in the sequence.

\paragraph{Assumptions.}

To express assumptions about committees and quorums,
we introduce a function \lstinline[language=Coq,basicstyle=\ttfamily\normalsize]{committee} that determines self-selected committees.
For example, the following statement says that
for any two quorums (i.e., subsets of size at least \lstinline[language=Coq,basicstyle=\ttfamily\normalsize]{tau})
of the committee for a given round-period-step triple,
there is an honest user who belongs to both quorums:
\begin{lstlisting}[language=Coq]
Definition quorum_honest_overlap_statement (tau:nat) :=
 forall (trace:seq GState) (r p s:nat) (q1 q2:{fset UserId}),
  q1 `<=` committee r p s -> #|q1| >= tau ->
  q2 `<=` committee r p s -> #|q2| >= tau ->
  exists (honest_voter : UserId), honest_voter \in q1 /\ honest_voter \in q2 /\
   honest_during_step (r,p,s) honest_voter trace.
\end{lstlisting}

Similarly, we capture that a block was certified in a period as follows (the value \texttt{3} indicates the third step, the \texttt{certvote} step, in period \texttt{p} and round \texttt{r}):
\begin{lstlisting}[language=Coq]
Definition certified_in_period (trace:seq GState) (tau r p:nat) (v:Value) :=
  exists (certvote_quorum:{fset UserId}),
  certvote_quorum `<=` committee r p 3 /\ #|certvote_quorum| >= tau /\
  forall (voter:UserId), voter \in certvote_quorum ->
   certvoted_in_path trace voter r p v.
\end{lstlisting}
This property is true for a trace
if there exists a quorum of users selected for cert-voting
who actually sent their votes in that trace for the given period
(via \lstinline[language=Coq,basicstyle=\ttfamily\normalsize]{certvoted_in_path}, which we omit).
This is without loss of generality since a corrupted user who did not send its cert-vote can be simulated by 
a corrupted user who sent its vote but the message is received by nobody.

\section{Asynchronous Safety}
\label{sec:safety}

The analysis of the protocol in the computational model
shows that the probability of forking is negligible~\cite{algorand-features-spec,Chen2019}.
In contrast, we specify and prove formally in the \emph{symbolic} model
with idealized cryptographic primitives 
that at most one block is certified in a round,
even in the face of adversary control over message delivery and corruption of users.
We call this property \emph{asynchronous safety}:
\begin{lstlisting}[language=Coq]
Theorem asynchronous_safety : forall (g0:GState) (trace:seq GState) (r:nat),
  state_before_round r g0 -> is_trace g0 trace ->
  forall (p1:nat) (v1:Value), certified_in_period trace r p1 v1 ->
  forall (p2:nat) (v2:Value), certified_in_period trace r p2 v2 ->
  v1 = v2.
\end{lstlisting}
Here, the first precondition \lstinline[language=Coq,basicstyle=\ttfamily\normalsize]{state_before_round r g0} states
that no user has taken any actions in round \lstinline[basicstyle=\ttfamily\normalsize]{r} in the initial global
state \lstinline[language=Coq,basicstyle=\ttfamily\normalsize]{g0},
and the second precondition \lstinline[basicstyle=\ttfamily\normalsize,breaklines=true]{is_trace g0 trace} states that
\lstinline[language=Coq,basicstyle=\ttfamily\normalsize]{trace} follows \lstinline[language=Coq,basicstyle=\ttfamily\normalsize]{~~>}$\!\!$ and starts in \lstinline[language=Coq,basicstyle=\ttfamily\normalsize]{g0}.

Note that it is possible to end up with block certifications from multiple periods of a round.
Specifically, during a network partition, which allows the adversary to delay messages,
this can happen if cert-vote messages are delayed enough
for some users to advance past the period where the first certification was produced.
However, these multiple certifications will all be for the same block.

\paragraph{Proof Outline.}
The proof of asynchronous safety proceeds by case-splitting on
whether the certifications are from the same period or different periods.
For\linebreak the first and easiest case, \lstinline[language=Coq,basicstyle=\ttfamily\normalsize]{p1 = p2}, we use quorum hypotheses
to establish\linebreak that there is an honest user that contributed a cert-vote to both certifications.
Then, we conclude by applying the lemma \lstinline[language=Coq,basicstyle=\ttfamily\normalsize]{no_two_certvotes_in_p},
which establishes that an honest user \lstinline[language=Coq,basicstyle=\ttfamily\normalsize]{u} cert-votes at most once in a period (proved by exhaustive analysis of possible transitions by an honest node):
\begin{lstlisting}[language=Coq]
Lemma no_two_certvotes_in_p : forall (g0:GState) (trace:seq GState) u (r p:nat),
 is_trace g0 trace ->
 forall idx1 v1, certvoted_in_path_at idx1 trace u r p v1 ->
   user_honest_at idx1 trace u ->
 forall idx2 v2, certvoted_in_path_at idx2 trace u r p v2 ->
   user_honest_at idx2 trace u -> idx1 = idx2 /\ v1 = v2.
\end{lstlisting}
The second case (\lstinline[language=Coq,basicstyle=\ttfamily\normalsize]{p1 <> p2}) uses an invariant which first holds in the period that produces the first certification, say, \lstinline[language=Coq,basicstyle=\ttfamily\normalsize]{p1} for \lstinline[language=Coq,basicstyle=\ttfamily\normalsize]{v1},
and then keeps holding for all periods of the round.
The invariant is that no step of the period produces a quorum of open next-votes,
and any quorum of value next-votes must be for \lstinline[language=Coq,basicstyle=\ttfamily\normalsize]{v1}. (Please refer to~\cite{AlgorandVerification} for the full definitions of predicates appearing in the lemma.)

\section{Conclusion}
\label{sec:conclusion}
\vspace{-5pt}
We presented a model in Coq of the Algorand consensus protocol
and outlined the specification and formal proof of its asynchronous safety.
The model and the proof open up many possibilities
for further formal verification of the protocol,
most directly of \emph{liveness} properties.
Our Coq development is available on GitHub~\cite{AlgorandVerification} and
contains around 2000 specification lines and 4000 proof lines.

%
%
%
\bibliographystyle{splncs04}
\bibliography{bib}
\end{document}